\documentclass[aps,prb,twocolumn,groupedaddress,showpacs]{revtex4}
\usepackage{graphicx}
\usepackage{latexsym}
\def\LSNiO{La$_{2-x}$Sr$_x$NiO$_{4+\delta}$}

\def\BSCCO{Bi$_2$Sr$_2$CaCu$_2$O$_{8+\delta}$}
\def\LNSCO{La$_{1.6-x}$Nd$_{0.4}$Sr$_x$CuO$_{4}$}

\begin{document}
\hyphenation{Ka-pi-tul-nik}

\title{Periodic Coherence Peak Height Modulations in Superconducting $\rm Bi_2Sr_2CaCu_2O_{8+\delta}$}

\author{A. Fang$^1$, C. Howald$^2$, N. Kaneko$^{3,4}$, M. Greven$^{1,3}$, and A. Kapitulnik$^{1,2}$ }

\altaffiliation {$^4$National Institute of Advanced Industrial Science and
Technology, Tsukuba Central 2-2, Tsukuba, Ibaraki 305-8568, Japan\\}
\affiliation{$^1$Department of Applied Physics, Stanford University, Stanford, CA 94305\\
$^2$Department of Physics, Stanford University, Stanford, CA 94305\\
$^3$ Stanford Synchrotron Radiation Laboratory, Stanford, CA
94309\\ }

\date{\today}

\begin{abstract}
In this paper we analyze, using scanning tunneling spectroscopy (STS),
the local density of electronic states (LDOS) in nearly
optimally doped {\BSCCO} in zero field.
We see both dispersive and non-dispersive spatial LDOS modulations
as a function of energy in our samples.  Moreover, a spatial map of the superconducting coherence 
peak heights shows the same structure as the low energy LDOS.  
This suggests that these non-dispersive LDOS modulations originate from an underlying charge-density modulation which interacts with superconductivity.  

\end{abstract}

\maketitle

\section{Introduction}
\vspace{-2mm}
The Scanning Tunneling Microscope (STM) has been an important tool in the study of
high-temperature superconductors since their discovery. Initially,
 a variety of gap sizes and structures
were found and introduced much controversy into the subject.
Later, a more coherent consensus among different groups emerged regarding
the surface properties of these high-$T_c$ materials.
To give a few examples, STM
studies revealed the nature of the superstructure in BSCCO
\cite{kirk}, the d-wave nature  of the gap and  its size
\cite{renner1}, the effect of local impurities, the emergence
of zero-bias anomalies \cite{yazdani,davis1,davis2}, and the
electronic structure of the core of vortices \cite{renner2,pan2}.
More recent measurements suggest that superconductivity may not be
homogeneous in high-$T_c$ superconductors. In particular, STM
measurements have found spatial variations of the gap size 
in YBCO \cite{lozanne} and BSCCO \cite{cren,howald1,lang}.

While gap inhomogeneities have been found to dominate the electronic 
structure at large measured bias, more ordered structures underlying 
the d-wave-like tunneling spectra have been found at lower energies. 
A current topic of great interest in high-$T_c$ superconductors 
is the presence of spatial modulations of the 
charge and spin densities.  Theoretical
\cite{zaanen,ek1,ek2,polkovnikov,zaanen1,white} and experimental
\cite{tranquada,lake,lake1,yslee,mitrovic,khaykovich,zhou,yamada}
evidence has been mounting in support of the possibility that
their ground state exhibits spin and charge density waves (SDW
and CDW), which may be primarily one-dimensional (i.e., stripes), or two-dimensional \cite{altman} with a characteristic wave vector in the Cu-O bond direction of $q_{\pi-0}=0.25(2\pi/a_0)$. In STM measurements such a  modulation was first seen by Hoffman {\it et al.}\cite{hoffman} in a magnetic field as a 2D checkboard pattern of LDOS, aligned with the Cu-O bonds, around vortex cores in slightly over-doped BSCCO with Ni impurities.  The reported modulations showed a checkerboard ordering vector of  $q_{\pi-0}=0.23(2\pi/a_0)$ 
extending to large distances when measured at bias energy $\sim$ 7 meV. Howald {\it et al.}\cite{howald2} shortly afterwards reported this same effect in zero field on similarly doped BSCCO crystals without intentional substitution impurities. The observed modulation with ordering wave vector $q_{\pi-0} = [0.25\pm0.03](2\pi/a_0)$ was found at all energies,  exhibiting features characteristic of a two-dimensional system of line objects.  Moreover, Howald {\it et al.} showed that the LDOS modulation manifests itself, for both positive and negative bias, as a shift of states from above to below the superconducting gap. The fact that a single energy scale (i.e. the gap) appears for both superconductivity and these modulations suggests that these two effects are closely related.

In subsequent studies at zero field, Hoffman {\it et al.}  \cite{hoffman2} and McElroy {\it et al.} \cite{mcelroy} measured the dispersion of the strongest Fourier peak along the $\pi-0$ (i.e. Cu-O) direction.  They asserted that it was consistent with what is expected from quasiparticle scattering interference \cite{wanglee}, in which a peak in the Fourier LDOS equals the momentum transfer wave vector of the incident and scattered waves.    In general, their data showed good agreement with photoemission results  (i.e. band structure results \cite{damascelli}) at large bias, but did not continue to disperse below $\sim$ 15 mV.  

To account for {\it all}  the available data in the full energy range, Vojta \cite{vojta}, Podolsky {\it et al.} \cite{podolsky} and  Kivelson {\it et al.} \cite{kivelson} have shown that both charge order and quasiparticle scattering effects  can occur in the presence of pinned fluctuating stripes.  
In particular, Podolsky {\it et al.} , using explicit calculations, showed that
for a system with incipient charge order the dispersion {\em for this particular effect} is very weak and less than expected by band-structure.   At low energy it converges to the ordering vector, rather than the vector corresponding to the nodal separation on the Fermi surface.
A similar conclusion was reached by Kivelson {\it et al.} who emphasized that at higher energies, for a relatively  clean material, quasiparticle scattering interference will show a strong signal that overwhelms the weak charge modulation.  At low energies, the minimum  energy required to overcome the finite superconducting gap means that only quasiparticles near the nodes participate and thus should give a wave vector for the interference which is larger than what was measured.
The discrepancy was taken as evidence that another phenomenon dominates at low energies.

In this paper, we present new data and analysis on this phenomenon, 
measured on near optimally doped samples.
Using the same apparatus with the same experimental conditions as before, 
we show data for two of our samples (from the same growth run): one which 
has a  (fixed) two-dimensional ordering wave vector $q_{\pi-0}=[0.25\pm0.03](2\pi/a_0)$ that dominates the dispersive signal over much of the energy range \cite{howald2}, and one which has a
stronger dispersive signal that begins to dominate from lower energies on up, yet still has a majority of spectral weight for the lowest energies at wave vector $q_{\pi-0}=[0.22\pm0.03](2\pi/a_0)$. 
We further show that for all the samples studied, the density of states at the gap (i.e. $dI/dV$ at $V = \Delta$)  exhibits modulations with a wave vector similar to the low-energy pattern.
Finally,  we present evidence for the interplay of these modulations and superconductivity by showing the strong supression of the large gap coherence peaks.

\section{Experiment}
\vspace{-2mm}

We performed measurements using a home-made cryogenic STM.  The STM measures 
differential conductance $G \equiv dI/dV$ which is proportional to the LDOS.  The samples
are near optimally doped (slightly over-doped)
single crystal $\rm Bi_2Sr_2CaCu_2O_{8+\delta}$ ($T_c \sim$ 86-87 K)\cite{bi2.1} grown by a 
floating-zone
method.  They are cleaved at room temperature in an ultra high vacuum
of better than  $1\times 10^{-9}$  Torr, revealing an atomically flat surface
between the BiO planes.  Then they are quickly lowered to
the cryogenic section at a temperature of 6-8 K, where cryopumping yields orders of 
magnitude better vacuum.  Typical data were taken with
a sample bias of $-200$ mV and a set point current of $-100$ pA, which establishes the
relatively arbitrary normalization for the LDOS.  
We also performed measurements 
with a sample bias of $+65$ mV and a setpoint current of $+25$ pA.

At each point on the surface, a spectrum ($dI/dV$ vs. sample voltage $V$) was taken.
The bias modulation for the spectra is 1 mV$_{rms}$.  This, in addition to the time 
constant of the lock-in amplifier used to record $dI/dV$, yields a total blurring
of the spectra of $\sim$ 3 mV.  
Although there are discrepancies when trying 
to fit a spectrum with a d-wave function, 
using the voltage, i.e. position, of the coherence peak maximum for the gap value $\Delta$
yields a reasonable fit, and is the method we will use throughout this paper.  
The coherence peak-heights map is made by evaluating $dI/dV$ at $V=\Delta$(at that location).
 All maps have been constructed 
for positive sample voltage, as this yields better
signal to noise. This observation is most likely related to the 
asymmetry of the conductance spectra which are common to all BSCCO 
STM studies \cite{renner3}.

\section{Results and Analysis}
\vspace{-2mm}
\subsection{Spatial Variation}

Spatial variations of the superconducting gap on the surface of BSCCO have been reported 
by several groups and their existance is now an established fact \cite{howald1,pan1,lang}.  Typically the size of the 
\begin{figure}[h]
\begin{center}
\includegraphics[width=1.0 \columnwidth]{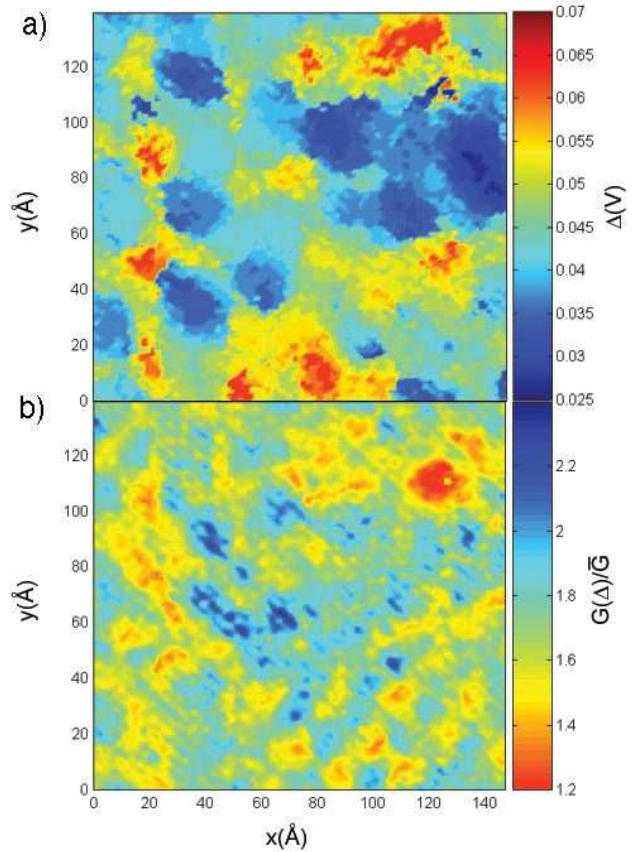}
\end{center}
\caption{a) Gap size $\Delta$ over a
140 \AA\ $\times$ 140 \AA\  area.  b) Coherence peak height (divided by the average conductance) with inverted colorscale.  Scan performed at $+65$ mV bias.} 
\label{maps}
\end{figure}
gap varies on a
 length scale of roughly 30 \AA\ as was described in detail elsewhere by Howald {\it et al} \cite{howald1}.  
The gaps 
 vary from small gap 
regions of $\sim$ 30 mV 
to large gap regions of $\sim$ 60 mV. 
The  regions of large gap usually develop from average gap background with the 
largest gap at their center \cite{howald1}. 
Fig.~\ref{maps}a shows the size of the gap 
mapped  over a typical surface. Patches about 30 \AA\ across of varying 
gap size are clearly visible.  The distribution of gap sizes in this 
view is depicted in Fig.~\ref{hsg}a, where the average gap is $\overline{\Delta}$
= 45.4 mV.  The smallest scale features 
reflect some variation with atomic resolution. In addition, the partial, 
nearly vertical lines show that there is some correlation between superstructure 
and the gap (Fig.~\ref{maps}a).

\begin{figure}[h]
\noindent
\includegraphics[width=.49 \columnwidth]{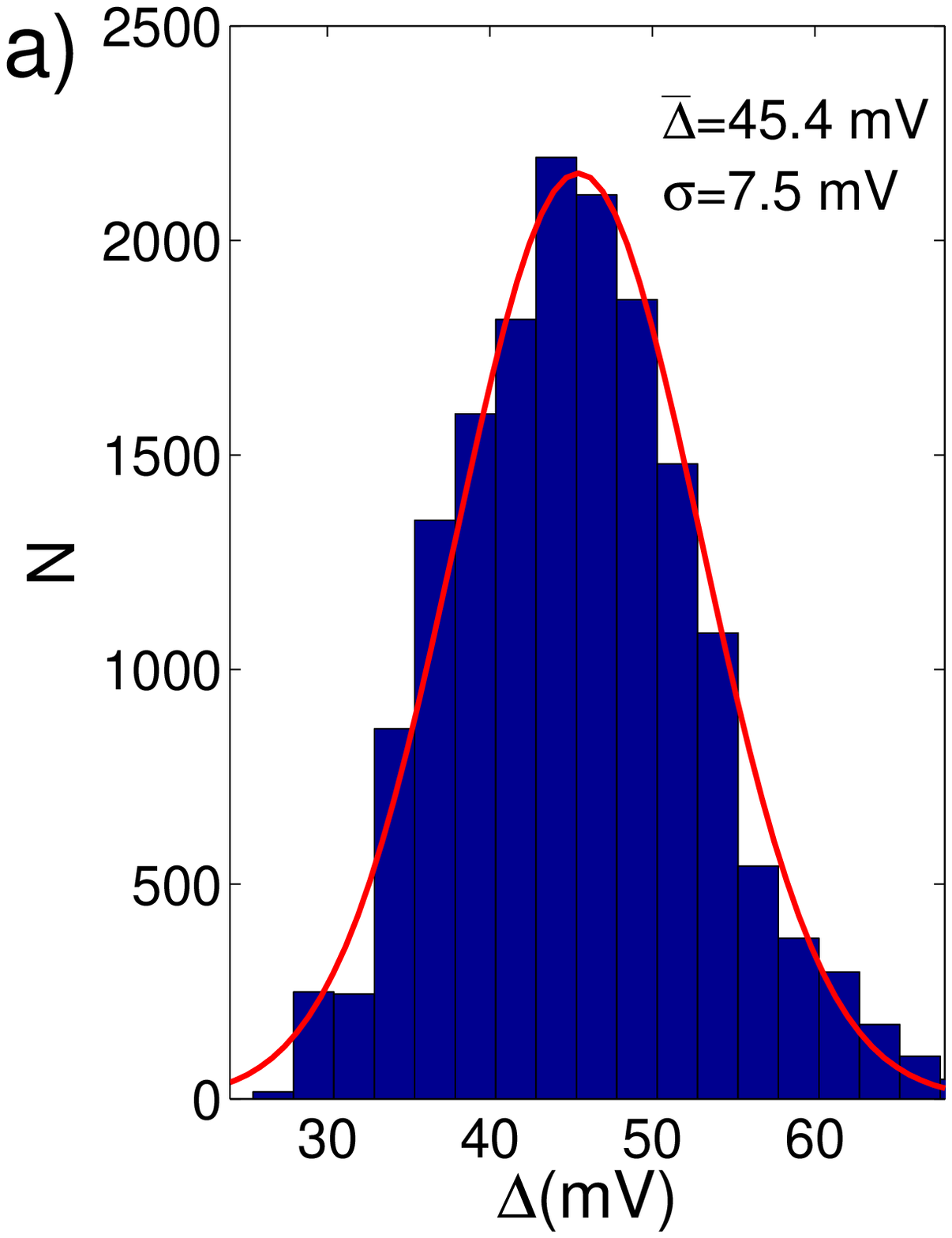} 
\includegraphics[width=.47 \columnwidth]{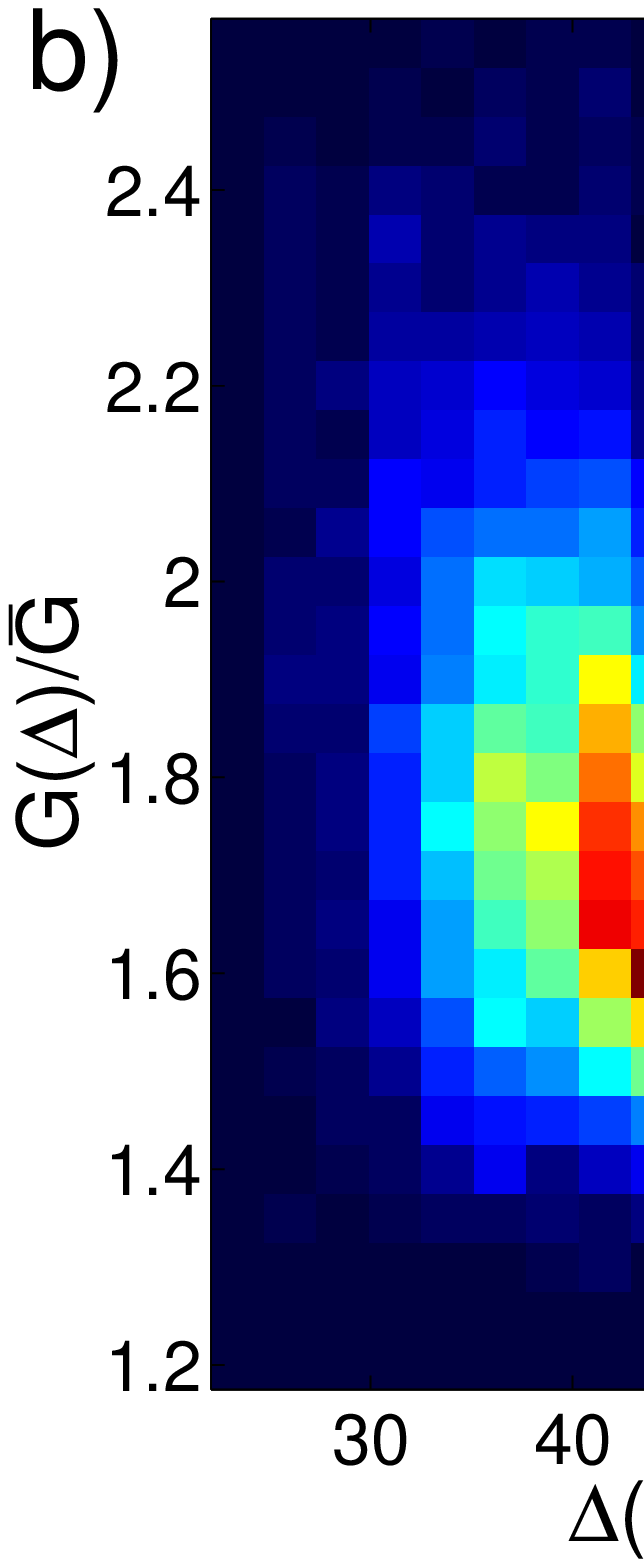}
\caption{a) Histogram of gap size distribution for the data in 
Fig.~\ref{maps}a.  b) Density plot 
of coherence peak height versus gap size. }
\label{hsg}
\end{figure}

An important feature of the gap-size distribution is that the height of the 
coherence peaks is varying as well.  There is some correlation between these two
effects: those spectra with the smallest gaps typically have
taller coherence peaks.  This anti-correlation is depicted in Fig.~\ref{hsg}b,
which shows the variation of the gap size and the
coherence peak height for the area in Fig.~\ref{maps}.
The colorscale of Fig.~\ref{maps}b is inverted
in order to demonstrate the anti-correlation shown in Fig.~\ref{hsg}b
(e.g.~regions with large gaps tend to have shorter coherence peaks).

While there is a correlation between gap size and coherence
peak height, it is not a simple one-to-one relationship.  
Thus the two maps can differ significantly. 
Aside from some high frequency noise,
the spatial variation in the peak height (Fig.~\ref{maps}b) is smoother, with no abrupt transitions 
between the regions. 
Next, we note the presence of an atomic corrugation which  
is stronger in one direction.  This 
is most likely due to a convolution with the tip shape, as this effect 
shows up in the topographic image as well. 
More importantly, it is clear that the peak-height 
shows an ordered structure (see e.g. the lower-right 
corner of the figure) in which the modulation amplitude can be as great as
$\sim$30$\%$ of the mean peak height in some areas.
 
By inspection of spectra from $-200$ mV to $+200$ mV, we found that normalization by the setpoint current at $+65$ mV makes the peak height map least sensitive to contributions from the 
superstructure as well as most of the low $k$-vector structure (which is due to gap size inhomogeneities).  Additionally, choosing a positive normalization voltage makes the spectra less dependent on the normal state background which is typically stronger on the negative bias side of the gap\cite{renner3}.
Thus taking the scan at $+65$ mV bias largely removes these features,
 as compared to
a peak height map taken with a more common setpoint voltage of $-200$ mV. 
This simple procedure allows the peak height structure to be seen in the real space image.  
A map of the current at $+65$ mV, for a spectroscopic scan taken
at $-200$ mV, shows little to no spectral weight above the noise
near $q = 0.25 (2\pi/a_0)$, and thus this procedure would not create the 
modulations we see.

\subsection{LDOS Modulations, Dispersion, and $G(\Delta)$}

\begin{figure}
\begin{flushleft}
\hspace*{0.2in}\includegraphics[width=0.6 \columnwidth]{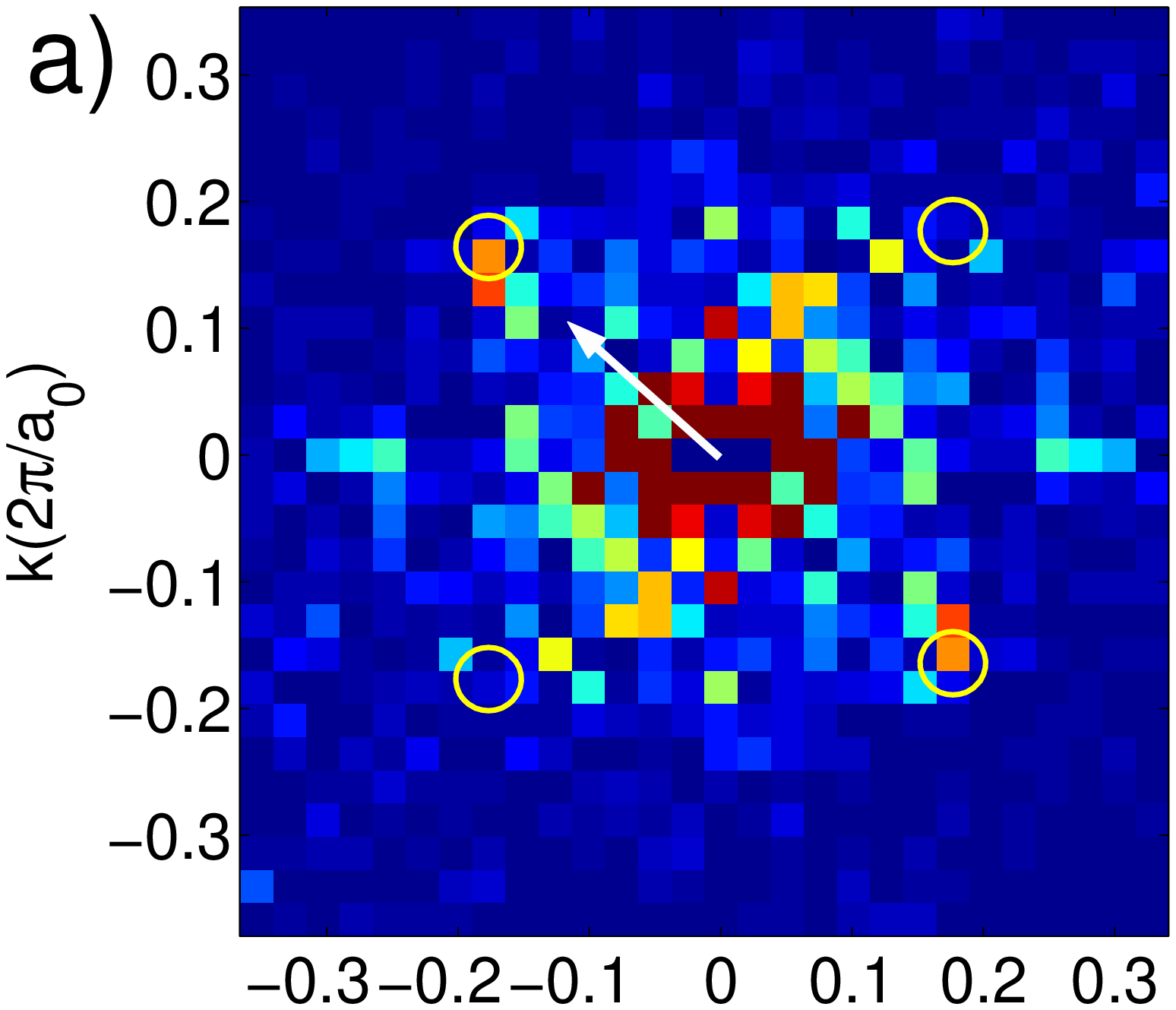}
\includegraphics[width=0.3 \columnwidth]{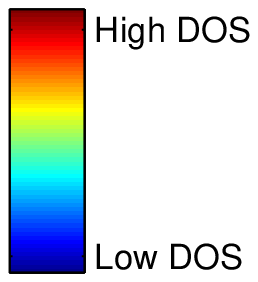}
\hspace*{0.2in}\includegraphics[width=0.6 \columnwidth]{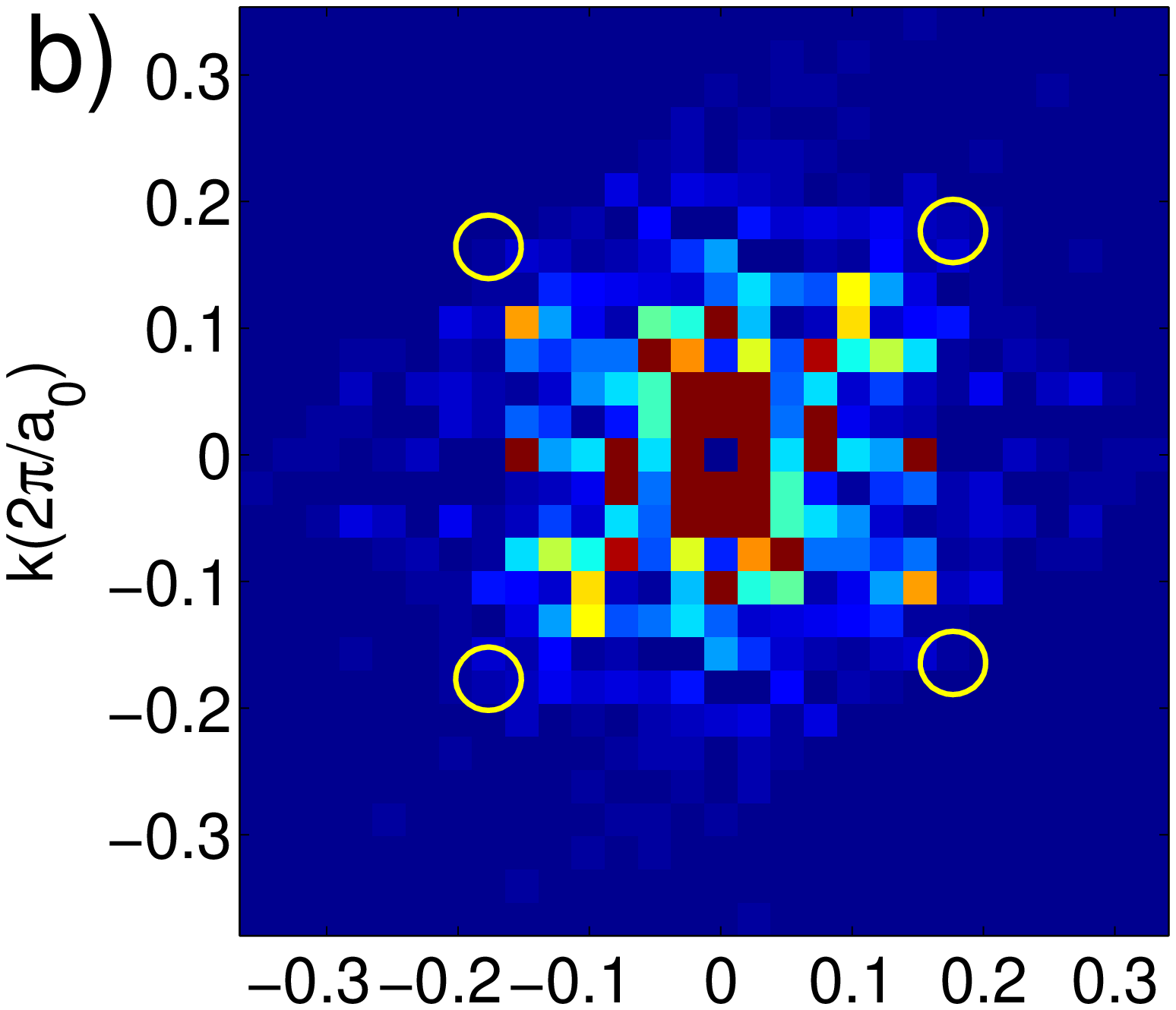}
\hspace*{0.2in}\includegraphics[width=0.6 \columnwidth]{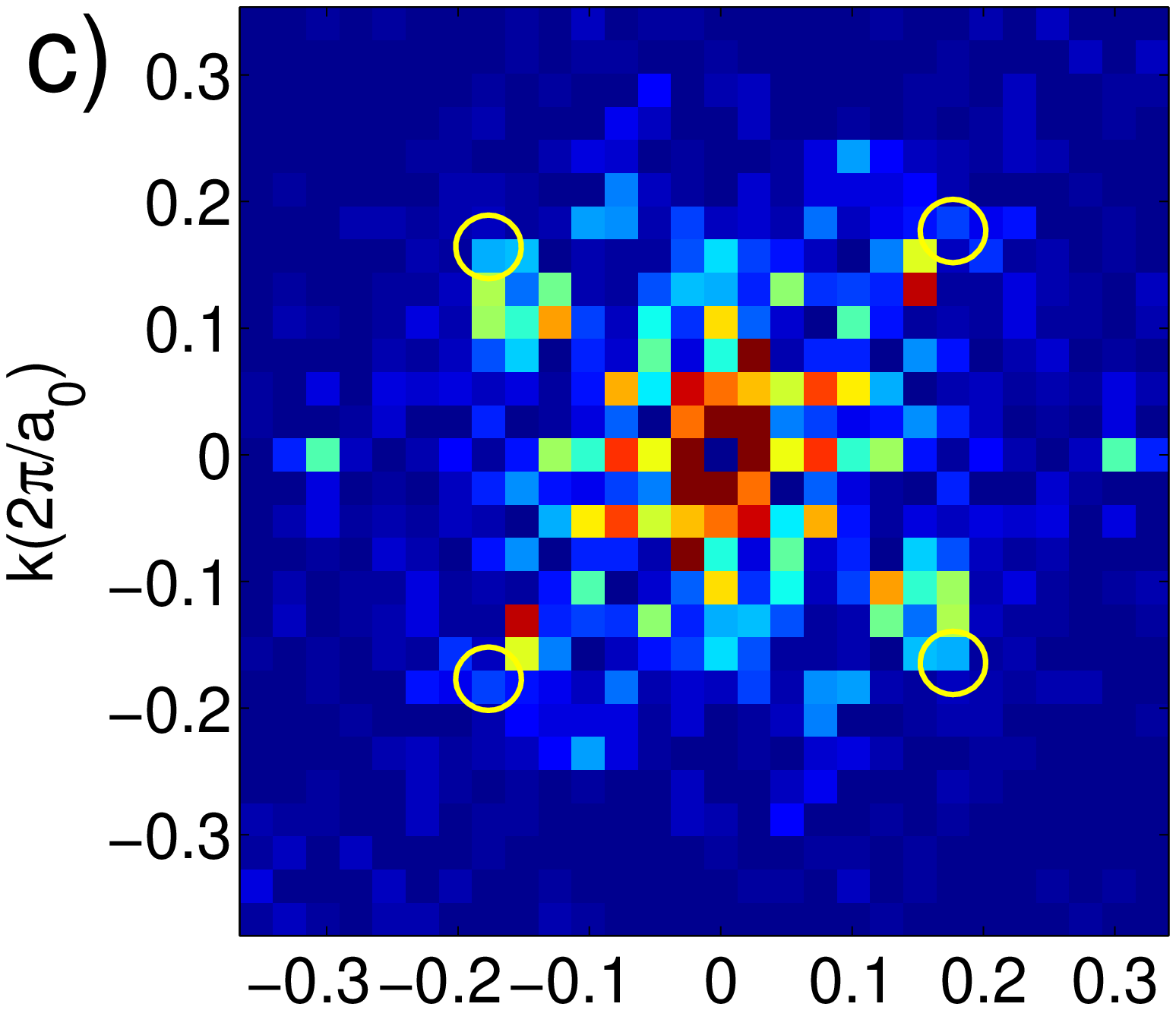}
\end{flushleft}
\begin{center}
\includegraphics[width=0.85 \columnwidth]{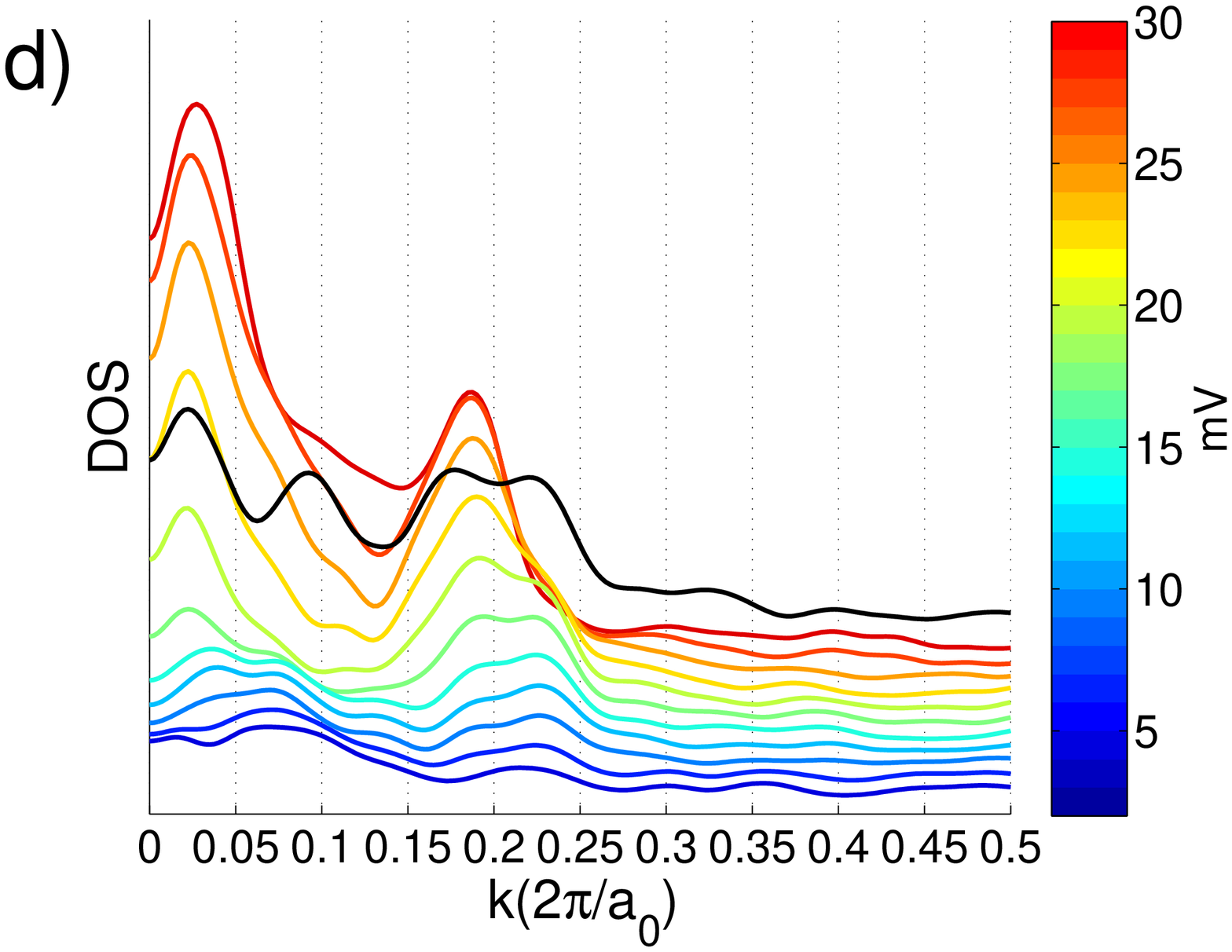}
\caption{a) FFT of LDOS $(dI/dV)$ at 10 mV. White arrow indicates direction of line scan. b) FFT of LDOS at 29 mV c) FFT of LDOS taken at the coherence peak maximum d) Dispersion relation of the charge modulation periodicity.  Black 
line is coherence peak maxima.  All lines shifted for clarity.}
\label{morefft}
\end{center}
\end{figure}

To look for LDOS modulations, one typically looks at the differential 
conductance $G \equiv dI/dV$
as a function of voltage ($V$) at each point on the sample.  In most 
cases modulations are more visible in Fourier space where the length 
scales of various features are better separated.  Fig.~\ref{morefft} 
shows a Fourier transform of the differential conductance of the area 
shown in Fig.~\ref{maps} for two different energies (10 mV and 29 mV) 
as well as for the gap, $\Delta$ (note that the majority of gaps are 30 mV $\leq \Delta \leq$ 60 mV 
for that region).  Circles are placed at (2$\pi /a_0$)($\pm$0.25,0) 
and (2$\pi /a_0$)(0,$\pm$0.25) as reference points.

\begin{figure}
\begin{flushleft}
\hspace*{0.2in}\includegraphics[width=0.615 \columnwidth]{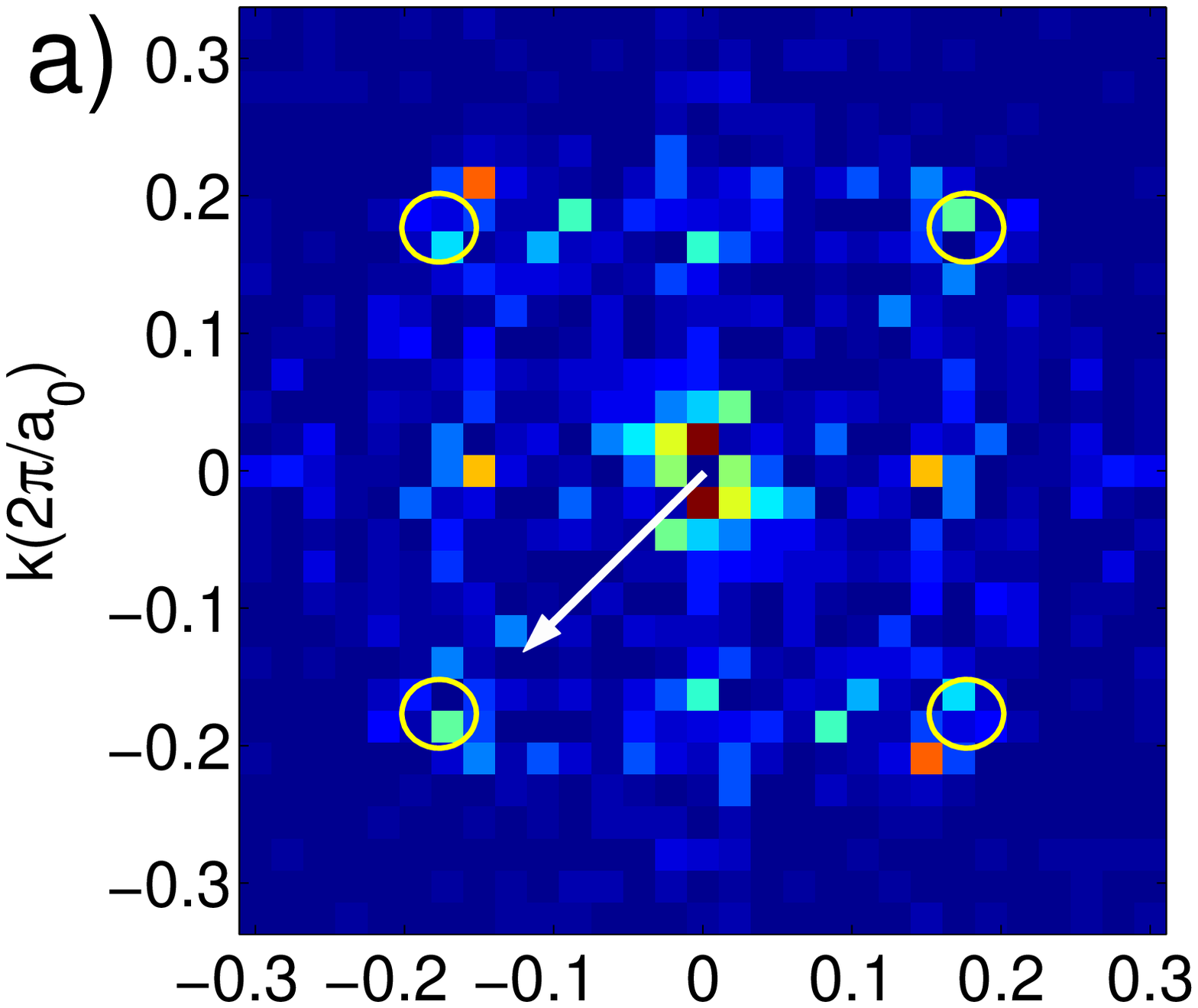}
\hspace*{0.2in}\includegraphics[width=0.615 \columnwidth]{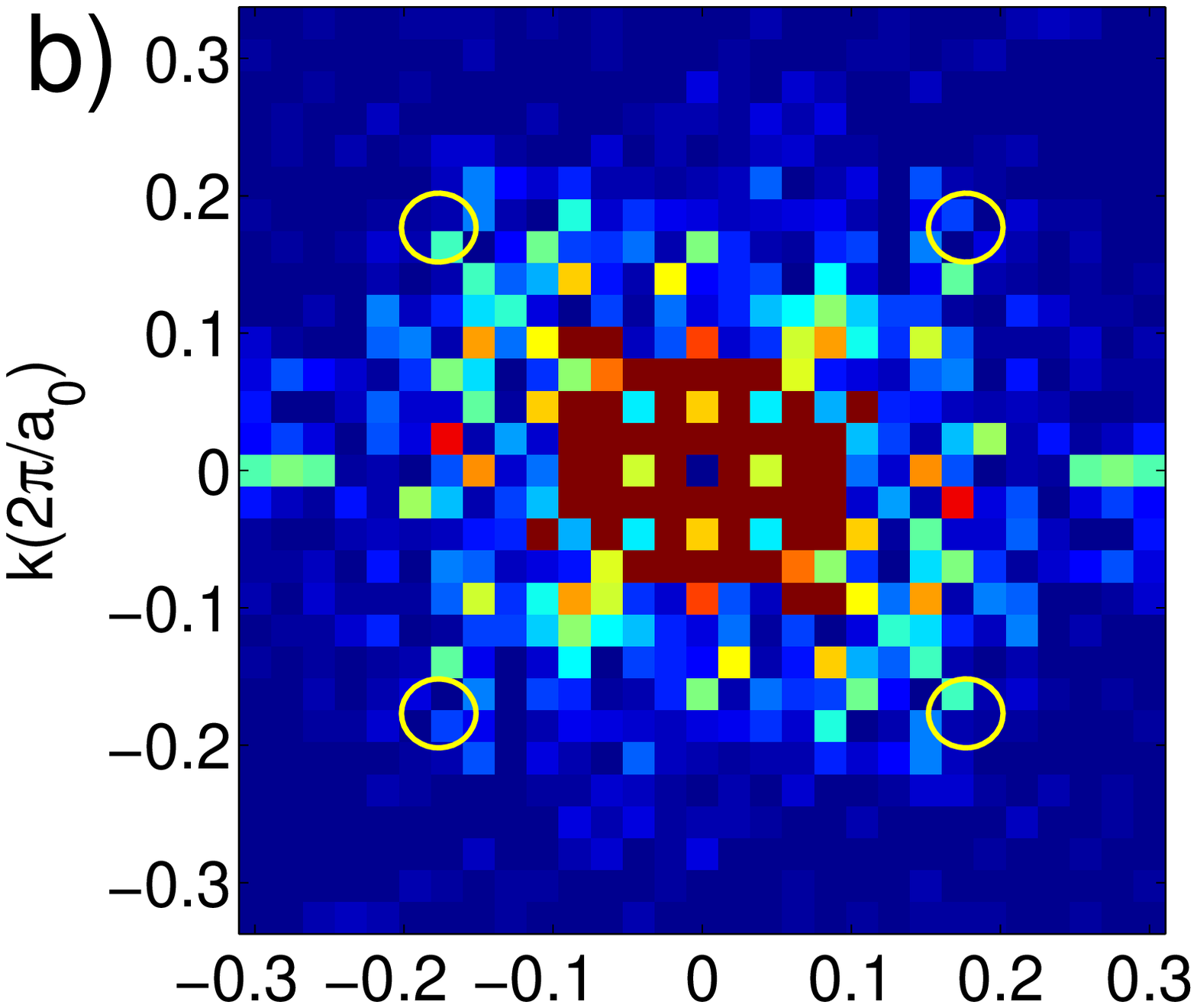}
\hspace*{0.2in}\includegraphics[width=0.615 \columnwidth]{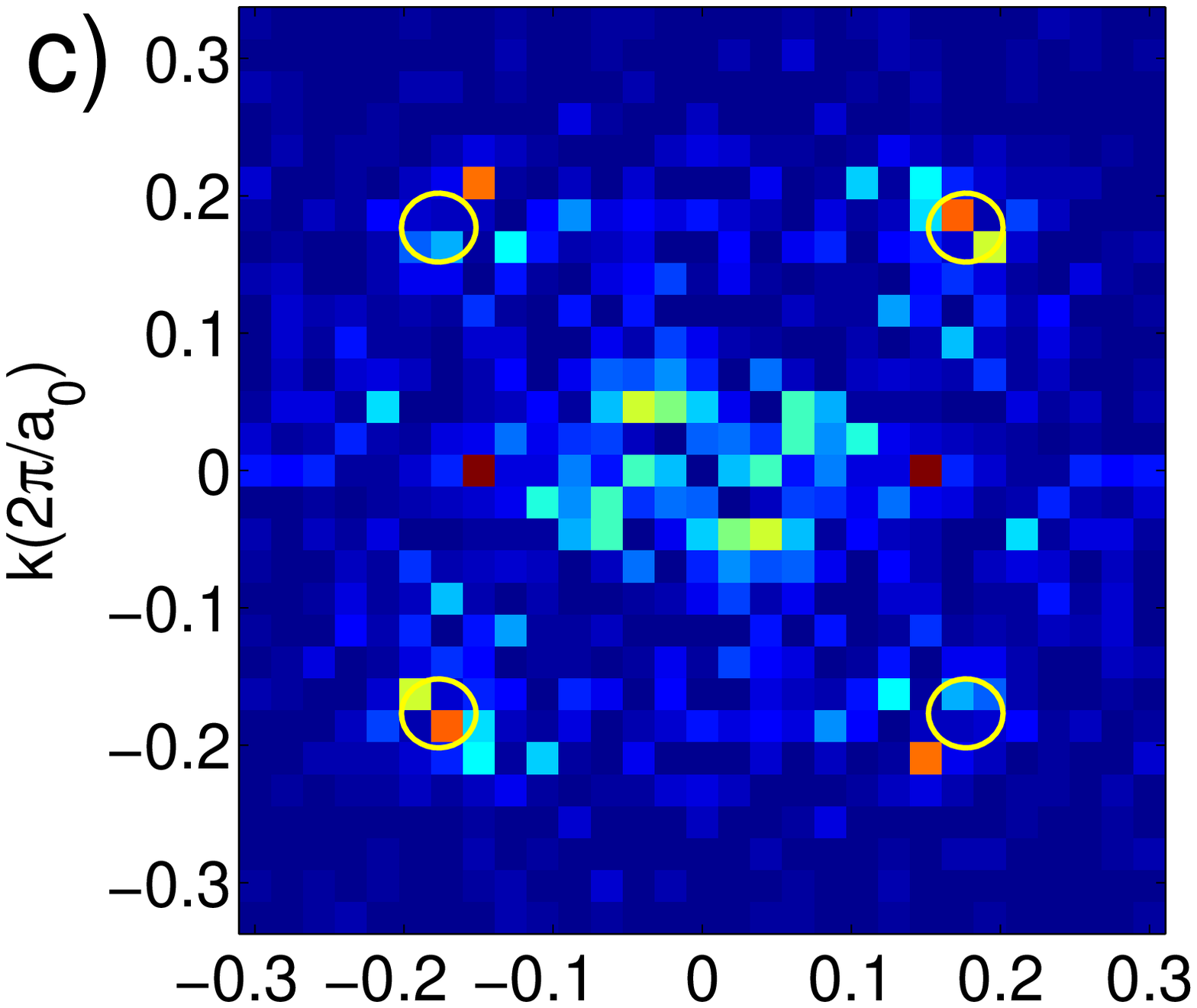}
\end{flushleft}
\begin{center}
\includegraphics[width=0.85 \columnwidth]{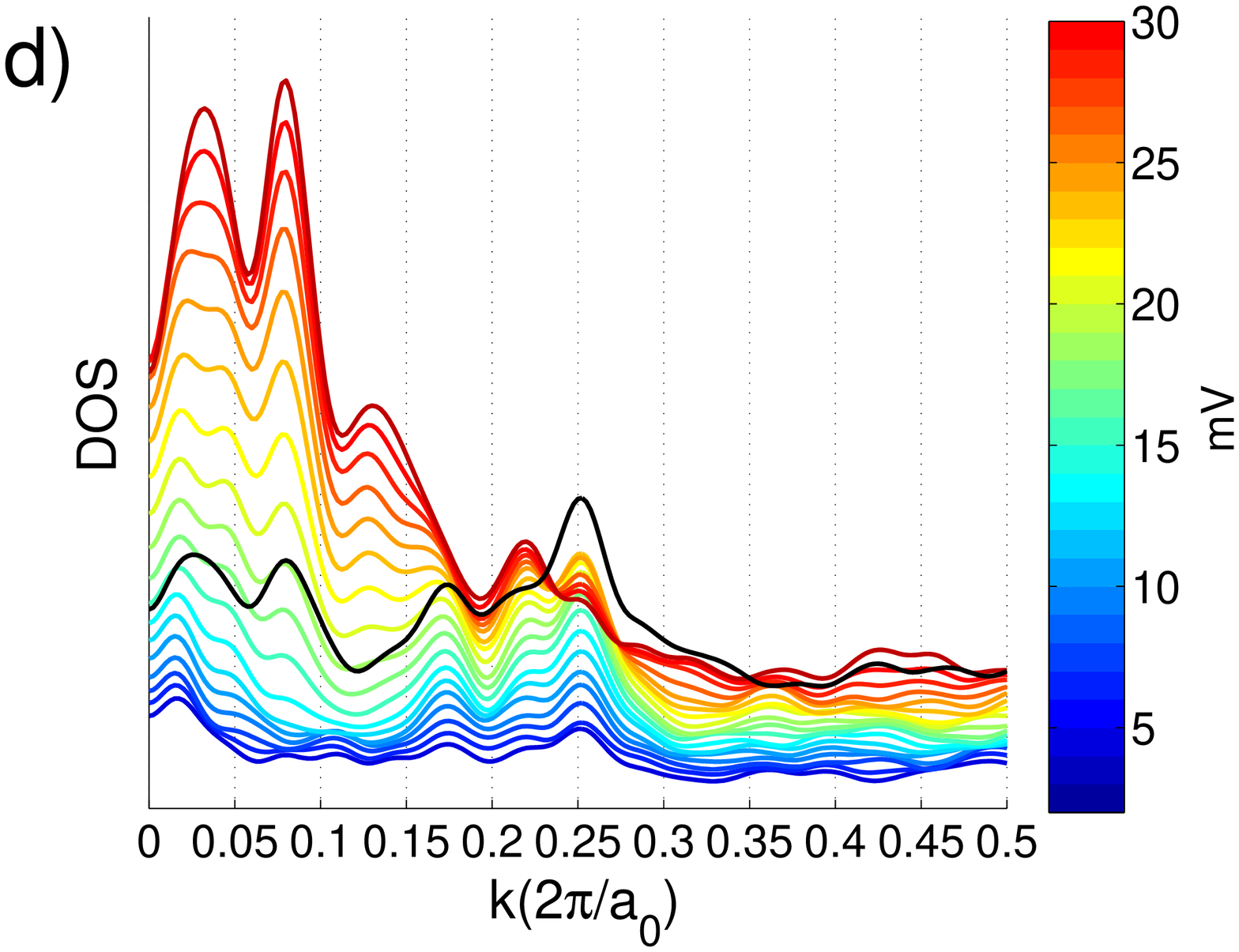}
\caption{Previously published data.
a) FFT of LDOS at 11 mV. b) FFT of LDOS at 28 mV 
c) FFT of LDOS taken at the coherence peak maximum with 65 mV normalization (see Section C.)
 d) Dispersion relation of the charge modulation periodicity.  Black 
line is normalized coherence peak maxima.}
\label{craigfft}
\end{center}
\end{figure}

As in previous results \cite{howald2} (Fig.~\ref{craigfft}), for low energies we see periodic density of 
states modulations at a periodicity close to  
$4a_0$.  The periodicity (as shown 
by the Fourier analysis) is $ [0.22 \pm 0.03](2\pi /a_0)$ for the data in Fig.~\ref{morefft}.  
At low energies ($\leq 15$ mV),
these modulations dominate (Fig.~\ref{morefft}a), but as one goes up in energy, the 
strength of the overall signal increases and moves to longer wavelengths 
in a very similar way to the results of McElroy {\it et al.} \cite{mcelroy}. 
This behavior is clearly seen in 
Fig.~\ref{morefft}b where we show a Fourier transform for a sample bias 
of 29 mV.  (The color scale on the Fourier maps has been adjusted to keep this 
signal within view.)
At higher energies, above $\sim$ 35 mV, the signal is lost 
in the noise due to the appearance of coherence peaks that vary with position on the sample. Although our experimental $k$-resolution limits precise quantitative statements we can make
about any dispersion, it is clear that overall, more spectral weight is appearing at lower $k$-vectors as the energy increases.  

To show this dispersive effect, we take line scans along the 
$\pi-0$ direction in Fourier space (Figs.~\ref{morefft}d and \ref{craigfft}d).  At each point along the line, we weight the neighboring Fourier points with a Gaussian filter of FWHM 1.6 pixels 
(which is not necessarily centered directly over a pixel).  The 
values are then squared, summed, and a square root is taken.  Finally, 
this value is normalized by the proximity of the line to the various
Fourier points.  The values reported in the line scans
are the total modulation amplitude in that region of Fourier space.
Due to pixelation effects and the size of our scans, the uncertainty in the peak positions, as well as our resolution, is  $\Delta q =  0.03(2\pi /a_0)$. 
However,  the normalization procedure and large filter width ensure that all of the features
(i.e. peaks) seen are distinct and not due to pixelation artifacts (e.g.~a broad peak will not be split into two by passing the line scan between pixels). 
 
We observe that at all the energies shown a (sometimes weak) signal always exists near the 
four-period wavevector,  in agreement with Howald {\it et al.}\cite{howald2}.  Additionally, there
exists a signal at a slightly lower $k$-vector.  This is most likely the dispersive
quasiparticle scattering interference peak since it becomes suppressed upon an
integration over energy, as shown for example in the above work\cite{howald2},  Fig. 9. If one follows the ``main peak" as suggested by Hoffman {\it et al.} (i.e. using a single peak to fit both features)
then dispersion is inferred at 
energies above $\sim$15 mV. 

If we consider the modulations in the peak heights (Fig.~\ref{maps}b)
 in Fourier space (Figs.~\ref{morefft}c,d), we find that it is very similar to the low energy modulation (Fig.~\ref{morefft}a).   The peak-heights map  has a periodic structure  close to four lattice spacings (more precisely $q_{\pi-0} = [0.22 \pm 0.03](2\pi/a_0)$) even though the entire contribution comes from energies above $\sim$ 30 mV (see Fig.~\ref{hsg}a).  
By looking at the LDOS based on gap size, we are sampling from a range 
of energies whose spectra contribute in such a way as to give a modulation at a higher $q$ than would 
be predicted by quasiparticle scattering for any of these energies alone.
The peak (or shoulder in the case of Fig.~\ref{craigfft}d) at slightly lower $q$  is likely to be from these quasiparticles at higher energies.
 
Finally, we note that to maximize the amplitude of the peaks, our line 
scan in Fig.~\ref{morefft}d is rotated by
$\sim$4 degrees with respect to the atomic positions.  This is not surprising
considering that our scan size is only over a few correlation lengths, and thus
defects can cause an overall rotation in the modulations.

\subsection{Interaction with Superconductivity}

\begin{figure}[h]
\begin{flushleft}
\hspace*{0.248in}\includegraphics[scale=0.4]{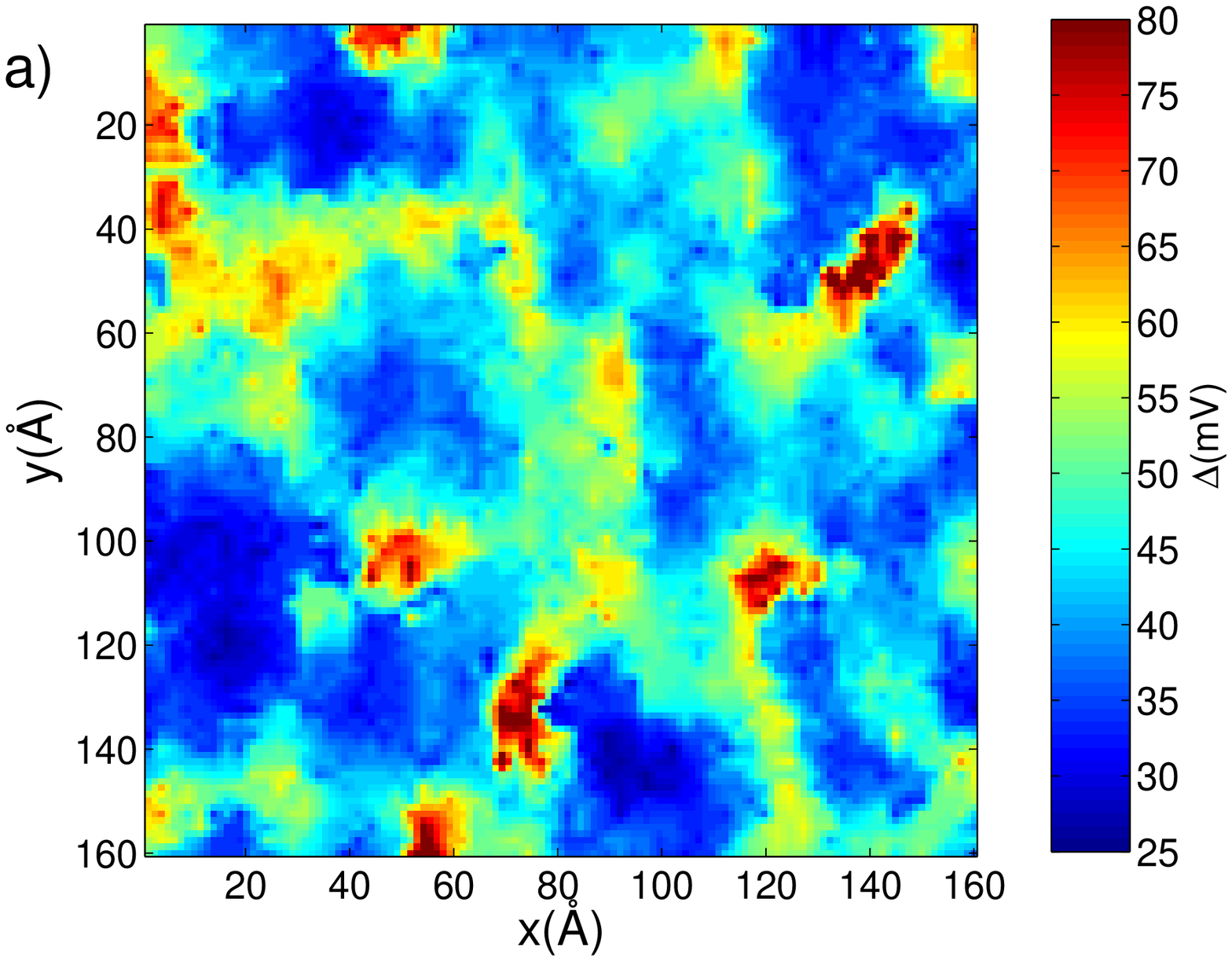}
\hspace*{0.26in}\includegraphics[scale=0.4]{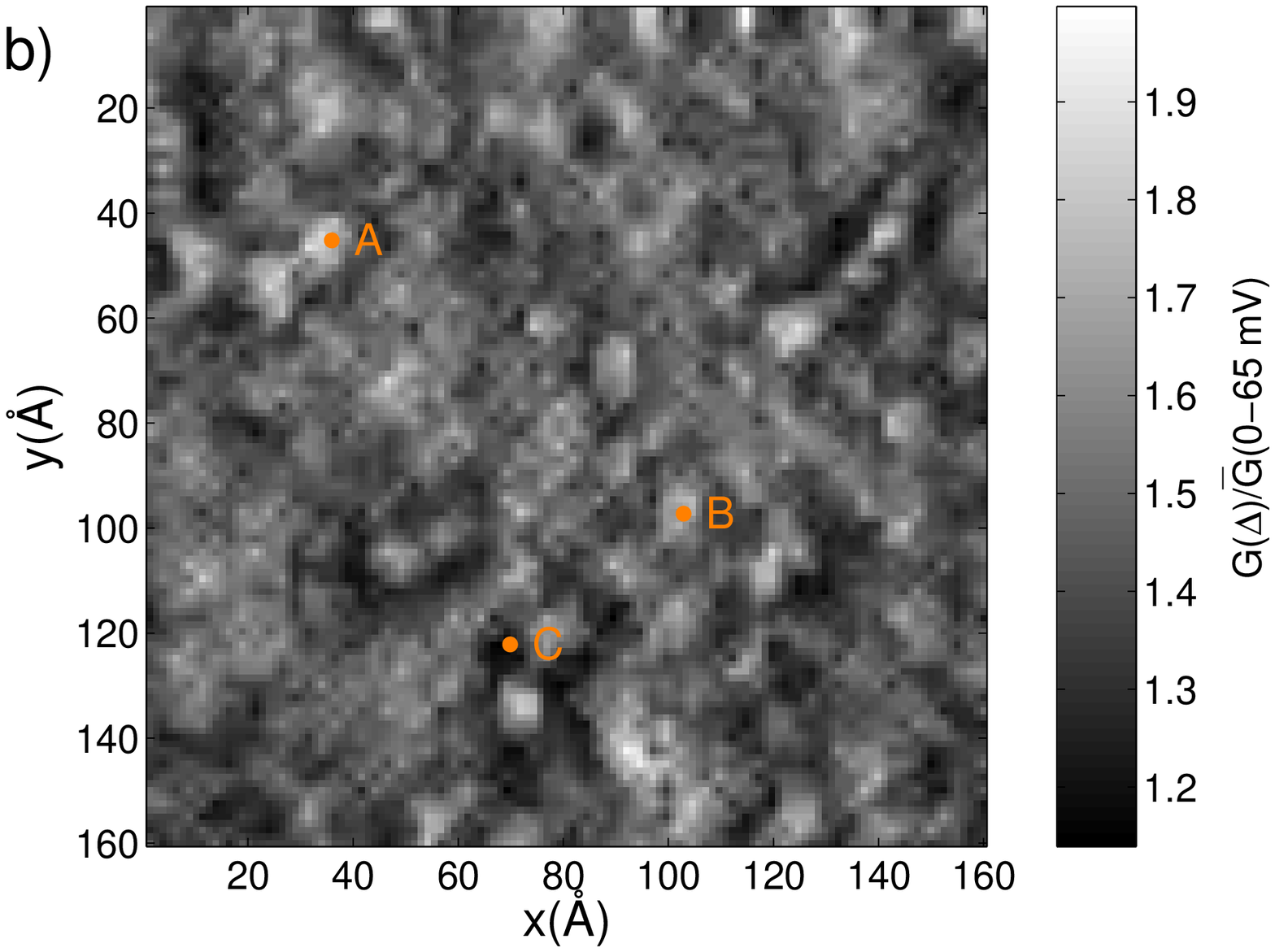}
\hspace*{0.266in}\includegraphics[scale=0.4]{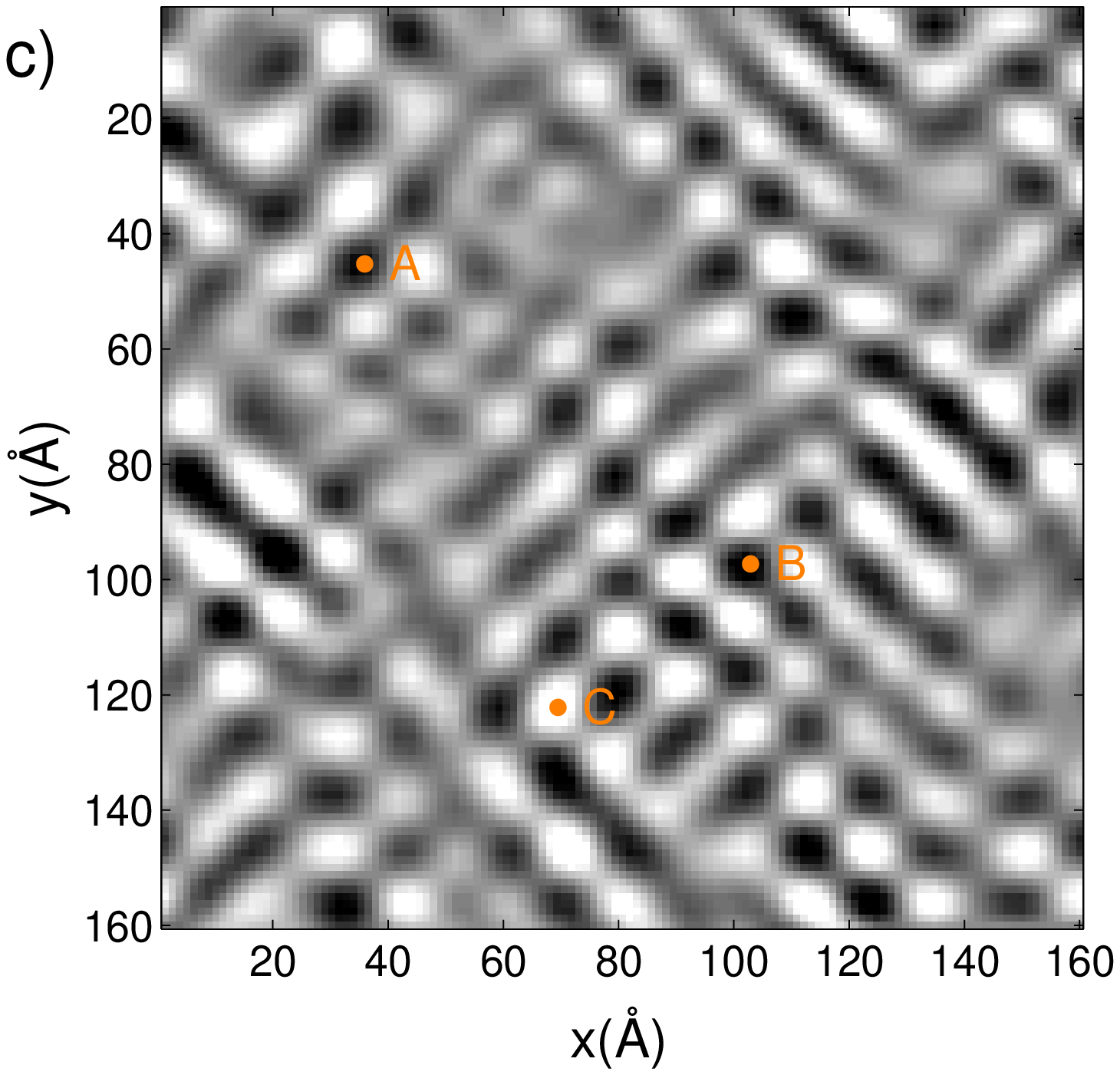}
\caption{a) Gap size distribution over a 160 \AA\ $\times$160 \AA\ area 
b) Coherence peak heights, normalized c) LDOS at 8 mV, un-normalized, Fourier filtered.  Equivalent points on Figs. b and c are marked by points A, B, C.}
 \label{mapscraig}
\end{flushleft}
\end{figure}

Some other interesting effects can also be seen when one compares 
the maps of the coherence peak heights and the gap size. 
Here we show previously published data \cite{howald2} that has been 
taken over a larger area (Fig.~\ref{mapscraig}).
A novel procedure we are using here is to normalize (i.e. divide) 
the individual spectra by the current at $+65$ mV.  (More precisely, we divide by the 
average conductance from 0 to $65$ mV.)
  This is similar to scanning over the 
area with this setpoint voltage of 65 mV (as in our previous figures).
We note that this procedure is only used to ``clean up" the image so
that certain features can be clearly seen in real space (note the lack of low frequency noise in Fig.~\ref{craigfft}c).  
As mentioned earlier, this procedure does not add any artificial 
modulations.  

By comparing Figs.~\ref{mapscraig}a and \ref{mapscraig}b (or Figs.~\ref{maps}a and \ref{maps}b),
one can see that the amplitude 
of the coherence peak DOS modulations is larger in the regions of large gap.  
In contrast, regions of small gap show modulations of reduced amplitude.  
Since there are only a few modulation crests and troughs within a particular
region of large or small gap, this effect is difficult to quantify, although
it can most easily be seen by following the regions of largest gap.  We note that the large amplitude modulations
are not simply due to the normalization procedure (regions of large gap are
renormalized to have higher LDOS due to the incomplete integration of the 
coherence peaks), since they are actually a larger
fraction of the mean coherence peak height.
  
Fig.~\ref{mapscraig}c shows the real space LDOS at 8 mV, after Fourier
filtering near the (2$\pi /a_0$)($\pm$0.25,0) and (2$\pi /a_0$)(0,$\pm$0.25)
points in reciprocal space.  We chose an energy where the signal from the underlying order dominates over the quasiparticle scattering interference signal.  
The filter is shaped like a circle, with a radius of 4 pixels in which
no Fourier weight is suppressed.  The edges then taper off like a Gaussian
with a width of one pixel.  This filter encompasses the vast majority of the spectral weight
in the region and thus does not favor one particular wave vector. The filtered image shows a dominant four-period modulation that is almost
checkerboard like, but with dislocations in the form of extra half-rows. 
By following the modulations, one can see a general correspondence
between the peaks of Fig.~\ref{mapscraig}b and the troughs of Fig.~\ref{mapscraig}c, and vice versa.  The points marked A, B, and C are examples of this out-of-phase relation.  (The features may not match up exactly due to noise or the slight contribution from quasiparticle scattering.) 
This suggests that where the low energy modulations have an increased LDOS, the coherence peaks are suppressed and vice versa.  In particular,
this suppression is stronger for the large gap coherence peaks.

\section{Discussion}
\vspace{-2mm}

It has been argued before that the competition between kinetic energy and Coulomb repulsion may lead to various forms of charge and spin ordered states.  In particular,  ``stripes"  have been predicted to occur  in doped antiferromagnets \cite{zaanen,schulz,ek1,subir}. 
The discovery \cite{hayd92,chen93,tran94a} of stripe
order in {\LSNiO} and soon after in {\LNSCO} \cite{tran95a} added considerable credibility
to the suggestion that stripe states form an important bridge between the
Mott insulator and the more metallic state which becomes Fermi-liquid-like at heavy doping. 
The theory is that at low doping, static stripes characterize a true broken symmetry state. However, upon increased doping the periodicity of the stripes decreases and coupling occurs.  This is roughly the region where superconducting order wins over stripe order.  Stripes will now exist in the superconducting phase in a dynamical sense, i.e.  stripe order is fluctuating. 
Therefore no broken symmetry is expected and thus to preserve the point symmetry of the underlying crystal, a fluctuating checkerboard is expected. 
STM, however, is a static probe and thus cannot detect any structure associated with fluctuating order unless something pins it
\cite{yazdani,davis1,pan2,howald2,pan1,lang,davis2,renner2,
hoffman,sprunger}. 
Indeed, the gap size inhomogeneities discussed above and other 
forms of chemical disorder \cite{eisaki04} are a natural source for pinning and can make stripes or checkerboards visible 
to STM in the form of LDOS modulations. 

Four-fold symmetrical order may also be a consequence of strong interactions on the square lattice.  When reduced to the low energy Plaquette Boson Fermion Model, the system shows a checkerboard structure due to the tendency of this model to locally prefer a four boson (an antiferromagnon triplet and a d-wave hole pair) state (i.e. doping of 1/8) \cite{altman,chen}.  In either of the above theories,  this underlying
order has a non- or weakly dispersive nature.
In the case of  ``striped" structures, weak dispersion is expected due to the finite 
size of the stripe domains and the interaction with the itinerant 
quasiparticles. 
\cite{kivelson,podolsky}

While the above two examples result in an (almost) ``fixed-$q$" order (i.e.~a true frozen charge density visible in the LDOS at all energies at an almost constant wave vector), the existence of quasiparticles (in the broad sense \cite{kivelson}) in the presence of weak disorder may also add  quasiparticle scattering interference effects\cite{wanglee,scalapino}. In this case, quasiparticles
of given energy scatter off an impurity.  The resulting interference between the original and 
scattered waves leads to variations of the local density of
states at wave vectors $\bf{q}=\bf k-\bf k^{\prime}$, where
$\bf k$ and $\bf k^{\prime}$ are the wave vectors of states with
energy $E=\epsilon(\bf
k)=\epsilon(\bf k^{\prime})$, as determined by the band structure,
$\epsilon(\bf k)$.  
Judging from measured band structure on BSCCO via photoemission \cite{damascelli}, quasiparticle scattering interference effects should be strongly dispersive, as is indeed seen in STM experiments for energies greater than $\sim$ 15 mV.

However, analysis of all recent experiments  \cite{kivelson} indicates that both evidence for a fixed-$q$  oscillation \cite{howald2} and quasiparticle scattering 
interference\cite{hoffman2,mcelroy} have been found 
experimentally.  Taking this point of view, it is clear that if  
there is an underlying order coexisting with the quasiparticle interference 
structure, one needs a way to separate out these effects. The main problem here
 is that the large contribution of the gap inhomogeneities and the strong 
 dispersion of the quasiparticle scattering interference cover up this underlying order.  Following the 
 ideas of Kivelson {\it et al.}, Howald {\it et al.} showed that integration of the Fourier space LDOS
 over a wide range of energies reduces the influence of any random or 
 dispersing features while at the same time enhances features that do 
 not disperse. 
  Another possibility is to look for the interaction of these features with
  another order parameter, such as superconductivity.
  We claim that the periodic structure 
  observed in the LDOS at the gap, $G(\Delta)$ in Fig.~\ref{maps}b, is exactly this effect. 

What we see is a return of features
from the low energy region (where saturation is observed in the plot of dispersion), suggesting that these features indeed exist and are
related to superconductivity. 
In addition, 
the large amplitude of these coherence peak modulations and the distinct peak near 
$q= 0.22 (2\pi/a_0)$, as shown 
in the line scan of the Fourier transform of the coherence peak heights (Figs.~\ref{morefft}d and \ref{craigfft}d), is suggestive of some kind of coherent contribution from
a non-dispersive feature. (Dispersive contributions would cause a peak in the line
scan to be broadened or suppressed, e.g. the peak/shoulder at lower $k$-vector.)

Our Fourier analysis from low energies up to the smallest gap sizes (where the noise
from inhomogeneities overwhelms our signal)
 supports the picture presented earlier of a non- or weakly 
dispersive feature near $q= 0.22 (2\pi/a_0)$ in addition to a dispersive feature at a lower $k$-vector.
We find that the large amplitude of this lower $k$-vector feature swamps out the non-dispersive feature at higher energies, but becomes
weaker as one goes down in energy, until at approximately 15 mV only the
non-dispersive signal remains, thus explaining the saturation. 
The additional line scan for the Fourier transform of coherence peak heights can be
seen as a way to remove the effects of gap size inhomogeneities to reveal that a structure
at $q\approx 0.22 (2\pi/a_0)$ still exists at higher energies.  It makes a similar
point as the spatial maps of the coherence peaks, namely, that by selectively
sampling from the (higher) energies related to superconductivity, the
low energy features reappear.

Finally, we note the correspondence of these (coherence peak height) modulations to 
the  superconductivity as determined by the size of the gap.  First we observe that the modulation structure that resides in the regions of larger gaps is the most pronounced.  These regions of large gap with low coherence peaks generally resemble slightly underdoped samples\cite{howald1}.  On the other hand, the modulation is suppressed and the coherence peak heights are more uniform in regions of small gap and tall coherence peaks.  Gaps  in these regions are more similar to gaps found in overdoped samples \cite{ozyuzer}; this is consistent with the notion that beyond optimal doping a more homogeneous  charge density exists as the system tends more towards a Fermi liquid state.
 
Such an observation does not necessarily point out a competition between charge-density modulation and superconductivity, but rather reinforces the idea that the two effects coexist at and below optimal doping.  One possible interpretation is that the fluctuating stripe/checkerboard phase exists  in all the regions below optimal doping, and as one moves further 
into overdoping (i.e. into regions of small gap) the modulations become diminished.
 Our observations therefore complement those of Vershinin {\it et al.}\cite{vershinin} who found similar patterns in the pseudogap regime of slightly underdoped $\rm Bi_2Sr_2CaCu_2O_{8 + \delta}$.  
As noted by Kivelson {\it et al.} \cite{kivelson}, the effect of quasiparticle scattering interference should disappear at temperatures above $T_c$, revealing the underlying order.
For our measurements at low temperature, in regions of very large gap with weak coherence peaks (similar to the pseudogap), charge ordering is indeed visible.  The two results therefore suggest that in the absence or suppression of superconductivity, charge-ordering may be the preferred phase.    

However, these ``pseudogap-like" regions are only about two superconducting coherence-lengths large, while the modulations (as seen in Fig.~\ref{mapscraig}c) persist over approximately seven periods \cite{hoffman,howald2}, suggesting that these modulations are more apparent in, but not exclusive to, these regions.  Moreover, both the modulation and superconductivity seem to coexist at low temperatures for all gap sizes except that in the small gap regions the peak-height modulation amplitude is suppressed as discussed above.  However, since the pseudogap is likely not a true phase transition, and in terms of both doping and temperature the system may be relatively far from a charge ordering critical point, ordering may be tenuous.  Upon lowering the temperature, the system undergoes a true phase transition into the superconducting phase where one expects the fixed-wavelength modulation phenomenon to be strongly pinned in the presence of disorder.
We see this effect as a non-dispersive signal which 
additionally manifests itself in the coherence peak heights of the tunneling spectra.

\section{Conclusion}
\vspace{-2mm}

The  observation of spatial modulations in a quantity that is related to the pair amplitude, i.e.~the coherence peak heights, 
points to the intimate 
relation between superconductivity and the charge modulation.    
This is strong evidence that the underlying 
modulation at $q \approx 0.22 (2\pi/a_0)$ which exists down to low energies, yet reappears at gap energies, is a separate effect from the quasiparticle scattering interference signal.
 Further analysis of the energy dependence suggests that the ``saturation" seen in plots of dispersion can be explained by the dominance of this non-dispersive signal at low energies.
Finally, the observation that these modulations suppress the coherence peaks most strongly in regions of large gap is consistent with a phase diagram that has an ordered phase towards under-doping and tends towards a more homogeneous charge distribution with increased doping.

 \bigskip

\noindent {\bf Acknowledgments:} We thank Assa Auerbach, Steven Kivelson, and John Tranquada for useful discussions. Work supported by the U. S. Department of Energy under contract No. DE-FG03-01ER45929.  
 The crystal growth was
supported by the U. S. Department of Energy under contracts No.
DE-FG03-99ER45773 and No. DE-AC03-76SF00515.


\begin{references}

\bibitem{kirk}
M.D. Kirk, J. Nogami, A.A. Baski, D.B. Mitzi, A. Kapitulnik, T.H.
Geballe, and C.F. Quate, Science, 242, 1673 (1988).

\bibitem{renner1}
C. Renner, O. Fischer, A.D. Kent, D.B. Mitzi, and A. Kapitulnik,
Physica B 194, 1689 (1994).

\bibitem{yazdani}
Ali Yazdani, C.M. Howald, C.P. Lutz, A. Kapitulnik, and D.M. Eigler,
Phys. Rev. Lett. 83, 176 (1999).

\bibitem{davis1}
E.W. Hudson, S.H. Pan, A.K. Gupta, K.W. Ng, and J.C. Davis, Science
285, 88 (1999).

\bibitem{davis2}
S.H. Pan, E.W. Hudson, K.M. Lang, H. Eisaki, S. Uchida, and J.C. Davis,
Nature, 403, 746 (2000).

\bibitem{renner2}
Ch. Renner, B. Revaz, K. Kadowaki, I. Maggio-Apprile, and O. Fischer,
Phys. Rev. Lett. 80, 3606 (1998).

\bibitem{pan2}
S.H. Pan, E.W. Hudson, A.K. Gupta, K.-W. Ng, H. Eisaki, S. Uchida,
and J.C. Davis, Phys. Rev. Lett. {\bf85}, 1536 (2000).

\bibitem{lozanne}
H.L. Edwards, D.J. Derro, A.L. Barr, J.T. Market, and A.L. de
Lozanne, Phys. Rev. Lett. 75, 1387 (1995).

\bibitem{cren}
T. Cren, D.Roditchev, W. Sacks, J.Klein, J.-B. Moussy, C. Deville-Cavellin, and M. Lagues, Phys. Rev. Lett. {\bf84}, 147 (2000)

\bibitem{howald1}
C. Howald, P. Fournier, and A. Kapitulnik, Phys. Rev. B {\bf64},
100504 (2001).

\bibitem{lang}
K. M. Lang, V. Madhavan, J. E. Hoffman, E. W. Hudson, H. Eisaki,
S. Uchida, and J. C. Davis, Nature {\bf415}, 412 (2002).

\bibitem{zaanen}
J. Zaanen and O. Gunnarsson, Phys. Rev. B {\bf40}, 7391 (1989).

\bibitem{ek1}
V.J. Emery, S.A. Kivelson, and H.Q. Lin, Phys. Rev. Lett. 64, 475 (1990).

\bibitem{ek2}
V.J. Emery and S.A. Kivelson, Physica C 209, 597 (1993); S.A.
Kivelson and V.J. Emery, in {\it Strongly Correlated Electronic
Materials}, Proceedings of the Los Alamos Symposium,
1993, ed. by K.S. Bedell, Z.Wang, D.E.Meltzer, A.V.Balatsky, and
E.Abrahams (Addison-Wesley Publishing Co, 1994) pg. 619.

\bibitem{polkovnikov}
A. Polkovnikov, S. Sachdev, M. Vojta, and E. Demler, in `Physical
Phenomena at High Magnetic Fields IV', October 19-25, 2001, Santa
Fe, New Mexico, International Journal of Modern Physics B, {\bf16} 3156 (2002).

\bibitem{zaanen1}
J. Zaanen and A.M. Oles, Annalen der Physik {\bf5}, 224 (1996).

\bibitem{white}
S.R. White and D.J. Scalapino,  Phys. Rev. Lett. {\bf81}, 3227
(1998).

\bibitem{tranquada}
J.M. Tranquada, J.D. Axe, N. Ichikawa, A.R. Moodenbaugh, Y.
Nakamura, and S. Uchida, Phys. Rev. Lett. {\bf78}, 338 (1997).

\bibitem{lake}
B. Lake, G. Aeppli, K.N. Clausen, D.F. McMorrow, K. Lefmann, N.E.
Hussey, N. Mangkorntong, M. Mohara, H. Takagi, T.E. Mason, and A.
Schroder, Science {\bf291}, 1759 (2001).

\bibitem{lake1}
B. Lake, G. Aeppli, K.N. Clausen, D.F. McMorrow, K. Lefmann, N.E.
Hussey, N. Mangkorntong, M. Mohara, H. Takagi, T.E. Mason, and A.
Schroder, Nature {\bf415}, 299 (2002).

\bibitem{yslee}
Y. S. Lee, R. J. Birgeneau, M. A. Kastner, Y. Endoh, S. Wakimoto,
K. Yamada, R. W. Erwin, S. H. Lee, and G. Shirane, Phys. Rev. B
{\bf60}, 3643 (1999).

\bibitem{mitrovic}
V. F. Mitrovi\'{c}, E. E. Sigmund, M. Eschrig, H. N. Bachman, W.
P. Halperin, A. P. Reyes, P. Kuhns, and W. G. Moulton, Nature
{\bf413}, 501 (2001).

\bibitem{khaykovich}
B. Khaykovich, Y. S. Lee, R. Erwin, S.-H. Lee, S. Wakimoto, K. J.
Thomas, M. A. Kastner, and R. J. Birgeneau,  Phys. Rev. B {\bf66}, 014528
 (2002).

\bibitem{zhou}
X. J. Zhou, T. Yoshida, S. A. Kellar, P. V. Bogdanov, E. D. Lu,
A. Lanzara, M. Nakamura, T. Noda, T. Kakeshita, H. Eisaki, S.
Uchida, A. Fujimori, Z. Hussain, and Z.-X. Shen,  Phys. Rev.
Lett. {\bf86}, 5578 (2001).

\bibitem{yamada}
K. Yamada, C.H. Lee, K. Kurahashi, J. Wada, S. Wakimoto, S. Ueki, H. Kimura, Y. Endoh, S. Hosoya, G. Shirane, R.J. Birgeneau, M. Greven, M.A. Kaster, Y.J. Kim, Phys. Rev. B.
{\bf57}, 6165 (1998)

\bibitem{altman}
E. Altman, A. Auerbach, Phys. Rev. B. {\bf65}, 104508 (2002).

\bibitem{hoffman}
J. E. Hoffman, E.W. Hudson, K.M. Lang, V. Madhavan, S.H. Pan, H.
Eisaki, S. Uchida, and J.C. Davis, Science {\bf295}, 466 (2002).

\bibitem {howald2}
C. Howald, H. Eisaki, N. Kaneko, M. Greven, and A. Kapitulnik,
Phys. Rev B {\bf67}, 014533 (2003).

\bibitem{hoffman2}
J. E. Hoffman, K. McElroy, D.-H. Lee, K. M. Lang, H. Eisaki, S.
Uchida, and J. C. Davis, Science {\bf297}, 1148 (2002).

\bibitem{mcelroy}
K. McElroy, R.W. Simmonds, J.E. Hoffman, D.-H. Lee, J. Orenstein, H. Eisaki, S. Uchida, and J.C. Davis, 
Nature, {\bf422}, 592, (2003).

\bibitem{wanglee}
Q. H. Wang and D. H. Lee, Phys. Rev. B {\bf67}, 020511 (2003).

\bibitem{damascelli}
A. Damascelli, Z. Hussain, and Z.-X. Shen, Rev. Mod. Phys ,{\bf75 no. 2}, 473 (2003).

\bibitem{vojta}
M. Vojta, Phys. Rev. B {\bf66}, 104505 (2002).

\bibitem{podolsky}
D. Podolsky, E. Demler, K. Damle, and B. I. Halperin,
Phys. Rev. B {\bf67}, 094514 (2003).

\bibitem{kivelson}
S. A. Kivelson, I.P. Bindloss, E. Fradkin, V. Oganesyan, J. M. Tranquada, A.
Kapitulnik, and C. Howald, Rev. Mod. Phys, {\bf75}, 1201 (2003).

\bibitem{bi2.1}
The actual composition is more likely $\rm Bi_{2.1}Sr_{1.9}CaCu_2O_{8+\delta}$.

\bibitem{renner3}
Ch. Renner and O. Fischer, Phys. Rev. B {\bf51}, 9208 (1995).

\bibitem{pan1}
S. H. Pan, J. P. O'Neal, R. L. Badzey, C. Chamon, H. Ding, J. R.
Engelbrecht, Z. Wang, H. Eisaki, S. Uchida, A. K. Gupta, K. W. Ng,
E. W. Hudson, K. M. Lang, and J. C. Davis, Nature {\bf413}, 282
(2001).

\bibitem{schulz}
H. J. Schulz, Phys. Rev. Lett. {\bf64}, 1445 (1990).

\bibitem{subir}
A. Polkovnikov, M. Vojta, and S. Sachdev, Phys. Rev. B {\bf65},
220509 (2002).

\bibitem{hayd92}
S.M. Hayden, G.H. Lander, J. Zaretsky, P.J. Brown, C. Stassis, P. Metcalf, and J.M. Honig,
Phys. Rev. Lett. {\bf68}, 1061 (1992).

\bibitem{chen93}
C.H. Chen, S.-W. Cheong, and A.S. Cooper, Phys. Rev. Lett. {\bf71}, 2461 (1993).

\bibitem{tran94a}
J.M. Tranquada, D. J. Buttrey, V. Sachan, and J.E Lorenzo, Phys. Rev. Lett.
{\bf73}, 1003 (1994).

\bibitem{tran95a}
J.M. Tranquada, J.E. Lorenzo, D.J. Buttrey, and V. Sachan, Phys. Rev. B {\bf52}, 3581 (1995).

\bibitem{sprunger}
P. T. Sprunger, L. Petersen, E. W. Plummer, E. Laegsgaard, and F.
Besenbacher, Science {\bf275}, 1764 (1997).

\bibitem{eisaki04}
H. Eisaki, N. Kaneko, D.L. Feng, A. Damascelli, P.K. Mang, K.M. Shen, Z.X. Shen, M. Greven,
Phys. Rev. B {\bf69}, 064512 (2004)

\bibitem{chen}
H.D. Chen, J.P. Hu, S. Capponi, E. Arrigoni and S.C. Zhang,
 Phys. Rev. Lett. {\bf89}, 137004-1 (2002).

\bibitem{scalapino}
L. Capriotti, D.J. Scalapino, and R.D. Sedgewick, Phys. Rev B {\bf68}, 014508 (2003).

\bibitem{ozyuzer}
L. Ozyuzer, J. F. Zasadzinski, K. E. Gray, C. Kendziora, and N. Miyakawa, Europhysics Lett. {\bf58}, 589 (2002).

\bibitem{vershinin}
M. Vershinin, S. Misra, S. Ono, Y. Abe, Y. Ando, A. Yazdani, Science, {\bf303}, 1995 (2004).


\end{references}
\end{document}